\documentclass[prb,aps,graphics,floats,showpacs,preprint]{revtex4}
\usepackage{graphicx}
\usepackage{amssymb}
\usepackage{amsmath}
\begin{document}

\title{Controlled nanostructures at La$_{0.7}$Sr$_{0.3}$MnO$_{3}$ thin film surfaces formed by STM lithography}

\author{LIU, Yun}
\affiliation{Department of Physics, Norwegian University of Science
and Technology, Trondheim 7491, Norway\\}

\author{ZHANG, Jia}
\affiliation{School of Mechanical Engineering, University of South
China, Hengyang, 421001, China\\}

\begin{abstract}
Nanoscale lithography on La$_{0.7}$Sr$_{0.3}$MnO$_{3}$ (LSMO) thin
film surfaces has been performed by scanning tunneling microscopy
under ambient conditions. From line-etching experiments we found
that the line depth is increasing in a step-wise fashion with
increasing bias voltage as well as with decreasing scan speed. On
average, the depth of the etched lines is an integral multiple of
the LSMO out-of-plane lattice constant about 0.4 nm. A minimum wall
thickness of 1.5 nm was obtained between etched lines. We have
utilized the ability to control the etched line depths to create
complicated inverse-pyramid nanostructure. Our work shows the
feasibility of using STM lithography to create controllable and
complex nanoscale structures in LSMO thin film.

\end{abstract}

\maketitle

\section{Introduction}

Perovskite manganites with chemical formula
La$_{1-x}$A$_{x}$MnO$_{3}$ (where A = Sr, Ba, Ca or Pb)
haveattracted large scientific interests due to their abundant but
complex physical properties, such as colossal magnetoresistance
(CMR) \cite{Helmot1993,Jin1994}, metal-insulator transition
\cite{Ramirez1997,Imada1998}, spin-, charge-, and orbital-ordering
\cite{Salamon2001, Coey2004}  as well as phase coexistence
\cite{Moreo1999,Varelogiannis2000,DagottoBook}. Some of the physical
properties have been proved to be strongly dependent on material
dimension down to nanometer length scale, for example, Takamura et
al \cite{Takamura2006} found that magnetic domain structures in
La$_{0.7}$Sr$_{0.3}$MnO$_{3}$(LSMO) thin films could be controlled
by varying shape (square, diamond, and circular) and size (diameters
ranging from $\sim$140 nm to 1$\mu$m) of patterned islands.
Therefore exploring those properties especially at the nanoscale
length plays a critical role in the study of manganites. Moreover,
owing to their CMR effect and high Curie temperature, the new
generation of spintronic devices is expected to exploit perovskite
manganites as major components. In particular, LSMO is showing a
Curie temperature as high as 370 K at intermediate hole doping
\cite{Urushibara1995}, and could serve as an promising candidate for
room temperature sensor and memory applications. Consequently
finding proper ways to create nanoscale structures on perovskite
manganite thin film surfaces are crucial not only for understanding
of basic phenomena at the nanoscale level but also for fabrication
of spintronic devices.

Scanning probe microscopy (SPM) based lithography \cite{Tseng2005}
has been a prevailing tool for controlled patterning on different
materials at the sub-100 nm length scale following the pioneering
work done by Dagata et al in 1990 \cite{Dagata1990}. Since then,
this technique has been successfully used to fabricate
nanostructures on graphite \cite{Albrecht1989,Penner1991}, metals
(Au \cite{Guo1992,Chang1994,Lebreton1996}, Ti \cite{Irmer1997}, Al
\cite{Snow1996}, and Cr \cite{Snow2002} ), semiconductors (Si
\cite{Lyo1991,Kobayashi1993}, SiO$_{2}$/Si \cite{Iwasaki2003}, and
GaAs \cite{Nagahara1990}) and perovskite oxides
(YBa$_{2}$Cu$_{3}$O$_{7}$ \cite{Heyvaert1992,Bertsche1998,Fan2000},
SrTiO$_{3}$ \cite{Pallecchi2002,Li2004}, SrRuO$_{3}$ \cite{You2007}
and La$_{0.8}$Ba$_{0.2}$MnO$_{3}$ \cite{Li2004,Li2005}), etc. In
general, the surface modification has been found to be strongly
dependent on etching parameters such as bias voltage, tunneling
current, scan speed, scan repetitions and ambient. For instance,
constant threshold voltage was observed for hole formation on gold
above a critical relative humidity \cite{Lebreton1996} as well as
for surface modification on graphite surface covered with water
\cite{Penner1991}, whereas the threshold voltage was found to be
dependant on the tunneling current for the hole formation on silicon
\cite{Kobayashi1993}.In line etching experiments on SrRuO$_{3}$ thin
film surfaces, the depth of etched lines was observed to increase
with increasing bias voltage and scan repetitions, while it
decreased with increasing scan speed \cite{You2007}. Li et al
\cite{Li2004,Li2005} conducted atomic force microscope (AFM)
lithography on La$_{0.8}$Ba$_{0.2}$MnO$_{3}$ films using conductive
tips (Si cantilever coated by Pt, Cr-Co and W$_{2}$C) in contact
mode. They found that only operation of negative sample bias voltage
created patterns with excellent controllability and reproducibility.
They also found that the pattern height increased with increasing
sample bias at first, then stabilized at different fixed voltages.
The smallest pattern width and interval were measured to be $\sim$50
nm by Pt or W$_{2}$C coated tips. The etching mechanisms in SPM
lithography are complicated and still under discussion. Among
others, widely proposed mechanisms are chemical reactions
\cite{Albrecht1989,Chang1994,Bertsche1998}, field evaporation
\cite{Lyo1991,Kobayashi1993,Heyvaert1992,Bertsche1998,Fan2000},
local heating \cite{Li1989}, electromigration \cite{Eigler1991}, and
combinations of these.

In the present work, we have exploited the feasibility of using
scanning tunneling microscope (STM) to create controlled nanoscale
structures at LSMO thin film surfaces. The influence of different
etching parameters has been addressed through line etching
experiments. The minimum wall thickness between etched lines has
been determined. Furthermore, we have successfully created
complicated nanostructures resembling inverse pyramid. Most
importantly, this work shows that nanostructures can be etched by
STM in LSMO in a highly controlled manner.

\section{Experimental}

LSMO thin films were deposited on TiO$_2$ terminated, single crystal
SrTiO$_{3}$ substratesby pulsed laser deposition (from TSST) using a
248 nm KrF excimer laser in an oxygen atmosphere with pressure 0.2
mBar. The laser energy is 50 mJ per pulse. The substrate temperature
was 850$^{\circ}$C during deposition. In situ RHEED monitoring of
the growth showed a continuous growth rate of 60 {\AA}/min. Films at
thickness 120 nm were used for the etching experiments. Structure
and surface characterization were performed using a high resolution
x-ray diffractometer (Bruker AXS D8) and an AFM (Digital Instruments
Nanoscope III), respectively.

The nanoscale lithography experiments were conducted in air at room
temperature with the AFM system operated in STM mode. Both
Platinum/Iridium (Pt/Ir) and pure iridium (Ir)tips were used. Images
of the etched patterns were recorded in situ after the lithographic
process with the same tips. Both imaging and lithography were
performed in constant current mode. Tunneling parameters for STM
imaging both before and after the etching process were a positive
tip (negative sample) bias voltage of 500 mV and a tunneling current
of 500 pA. The same tunneling parameters combined with sharp step
edges seen in the STM images were used as an indication of similar
tip quality before and after etching. During the lithographic
process, the tip movement was defined by a Nanoscript$^{TM}$
programme. All the etching experiments were conducted at large
positive tip bias voltages ($\geqslant$2.8 V). An ac current
component with frequency 50Hz and amplitude 40-60 pA was imposed to
60 pA setpoint tunneling current during STM lithography to prevent
material build-up during etching \cite{You2007}. In addition, by
adjusting the integral and proportional gains in the feedback loop,
the vertical oscillation amplitude of the STM tip could be
controlled within 1 nm. This method not only improved the quality of
the lithographic structures but also hindered possible damage to the
The tip by material build-up on the surface.

\begin{figure}[h!]
\begin{center}
  \includegraphics[width=9cm]{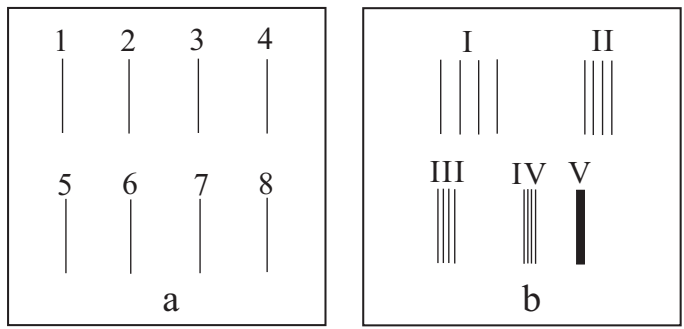}\\
  \caption{Schematics of (a) eight-line pattern and (b) pattern for wall thickness determination.
  }\label{fig:lineetching}
\end{center}
\end{figure}

\begin{figure}[h!]
\begin{center}
  \includegraphics[width=9cm]{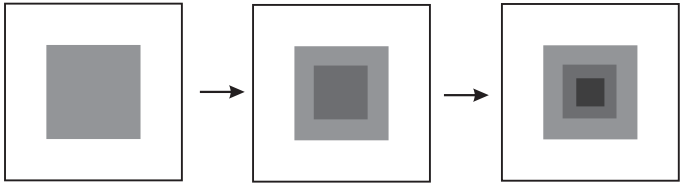}\\
  \caption{Schematics representation of etching procedures for (a) inverse-pyramid and (b) pyramid nanostructure formation.}
  \label{fig:pyramidetching}
\end{center}
\end{figure}

Figure \ref{fig:lineetching}(a) illustrates the eight-line pattern
etched duringthe line-etching experiments. The eight-line patterns
were created using bias voltages ranging from 2.8 V to 3.6 V, a
tunneling current of 60 pA, scan speeds from 500 nm/s to 1500 nm/s
and 100 scan repetitions per line. The length of each etched line
was 100 nm. The upper left line was etched first and the lower right
last, the sequence is indicated by the numbers in figure 1(a).
Figure \ref{fig:lineetching}(b) shows the patterns etched to
determine the thicknesses of walls formed between etched lines. The
etching parameters are bias voltages of 3.2 V, 3.4 V and 3.6 V, a
tunneling current of 60 pA, scan speeds of 500 nm/s and 1000 nm/s,
and 100 scan repetitions per line. From these experiments a minimum
wall thickness was determined below with areas could be formed by
numerous line etchings. Schematic three-step process of creating
inverse-pyramid structures is shown in \ref{fig:pyramidetching}.
First, the largest square was etched on the LSMO surface (left
panel), then the second largest square was created inside the
largest one (middle panel), and finally the smallest one likewise
(right panel). The same etching parameters were used through all
three steps. Bias voltages of 3.2 V and 3.4 V were selected to
create inverse-pyramid structure while other parameters were fixed
at scan speed 1000 nm/s, tunneling current 60 pA, and 100 scans per
line.

\section{Results and discussion}

From the x-ray diffraction measurements we determined the out of
plane $c$ parameterof the LSMO films to be $\sim$3.85 {\AA},
calculated from the (001), (002) and (003) diffraction peaks. This
value is slightly lower than the bulk value of 3.88 {\AA}, and
corresponds to an in-plane tensile strain associated with epitaxial
growth on the larger unit cell of SrTiO$_{3}$. The surface of the
films displayed the step-and-terrace topography, with atomically
flat plateaus, separated by steps of integral unit cell step
heights. The RMS roughness was measured to be approximately 0.14 nm.
Four point RT measurements showed metallic behavior (at room
temperature R = 5$\times$10$^{-3}$ Ohm $\cdot$ cm).

\begin{figure}[h!]
\begin{center}
  \includegraphics[width=9cm]{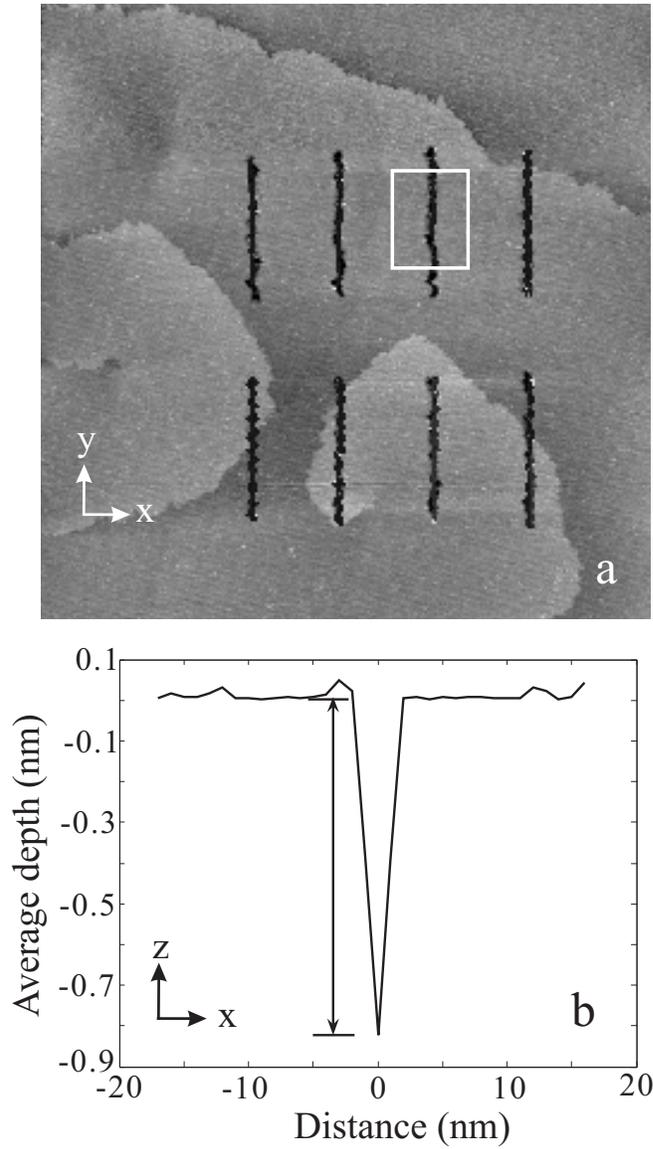}\\
  \caption{(a) STM image (500nm$\times$500nm) of eight etched lines from the line etching experiments.
  Each line was 100nm long and etched at bias voltage 3.4V,
  tunneling current 60pA, scan speed 1000nm/s and 100 scan repetitions per line.
  (b) section profile from the area confined by the dashed-line box. Each data point represents
  the average depths measured along the y-direction inside the marked area for each x-position.
  }\label{fig:depthdefinition}
\end{center}
\end{figure}

Figure \ref{fig:depthdefinition}(a) shows an STM image of an
eight-line pattern etched at a bias voltage of 3.4 V, scan speed
1000 nm/s, tunneling current 60pA and 100 scans per line. The depth
profile across an etched line is displayed in
\ref{fig:depthdefinition}(b). The average depth at each x-position
is determined from the depths measured along the y-direction inside
the marked area with length 70 nm along y. The two end parts of each
line were removed in the depth determination because there is a time
lag when the tip is changing direction at the ends of each line,
leading to deeper etching. Additionally, we define a line to be
successfully etched when the surface modification is continuous for
at least 70 nm with a minimum average depth of half a unit cell,
that is 0.2 nm. Only successful lines were analyzed.

\begin{figure}[h!]
\begin{center}
  \includegraphics[width=9cm]{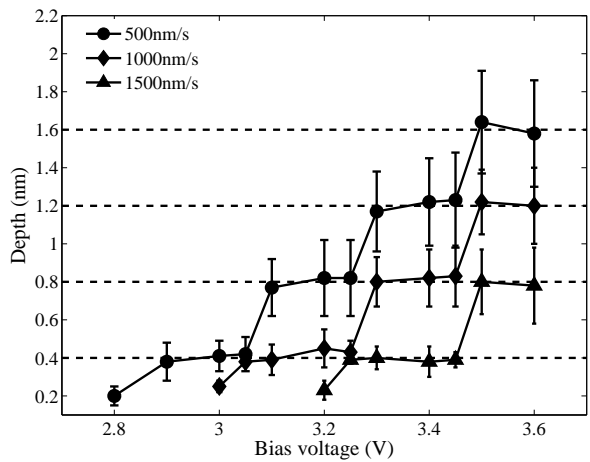}\\
  \caption{Dependence of etched line depth on bias voltage and scanspeed. The scan repetition for each line is
  100 and the setpoint tunneling current is 60 pA. 40 lines were calculated for each set of voltage and scan speed.
  The error bars are standard deviation and the solid lines are guides to the eye.}\label{}
  \label{fig:depthaVsvoltage}
\end{center}
\end{figure}

The depth determined from the line etching experiments are
summarized in figure \ref{fig:depthaVsvoltage}. We plot the line
depth as a function of bias voltage using 100 scan repetitions per
line for scan speeds of 500, 1000 and 1500 nm/s. Generally, the line
depth increases with increasing bias voltage in a stepwise fashion
with a step size of about 0.4 nm. Inversely, the depth increases
with decreasing scan speed. It should be mentioned that there are
three data points around 0.2 nm with very short errorbars. This is
partly a consequence of the chosen definition of successful etching.
As is evident from comparing the etching results at the three
different scan speeds, the longer line etching time corresponds to
slower scan speed. This finding is supported by another set of
experiments (not shown) revealing that a larger number of scan
repetitions per line leads to larger line depth. Thus, the STM
lithography is a time dependent process. These results are in
agreement with earlier studies of STM etching of
SrRuO$_{3}$\cite{You2007}, where also the line depth was found to
increase with the numbers of scan repetitions per line as well as
decreasing scan speed.

The most significant observation in figure \ref{fig:depthaVsvoltage}
is the change in the average line depth in steps of $\sim$0.4 nm,
that is, the average line depths are distributed around four levels:
0.4 nm, 0.8 nm, 1.2 nm and 1.6 nm, which are integral multiples of
the out-of-plane lattice constant ($\sim$0.4 nm) of the LSMO unit
cell. To our knowledge, this is the first work by STM lithography to
manipulate material surfaces at step of unit cell. The only
exception from this behavior is the observed etching close to
threshold where a depth of 0.2 nm  corresponding to half a unit cell
is obtained. The reason for the sub unit cell etching at the
outermost surface region of LSMO thin films may be related to the
broken symmetry at the surface. However, we could not observe half
unit cell steps on the LSMO thin films surfaces.  The etching
results at bias voltages of 3.05 V, 3.25 V and 3.45 V indicate that
there are critical bias voltages at which a transition between
different etching depths occurs. These findings suggest that it is
possible to control the depth of etched structure in LSMO with unit
cell precision.

At a given scan speed, we could identify threshold voltages below
which no etching occurs. For STM etching on
SrRuO$_{3}$\cite{You2007}, it is also found that the threshold
voltage depends on scan speed. In AFM lithography on LBMO films
using Pt-coated tip, at scan speed of 500 nm/s, Li et al
\cite{Li2005} reported a threshold voltage of $\sim$3 V at positive
tip bias, which is very similar to our results showing a threshold
voltage of 2.8 V at the same scan speed. They also observed that the
depth of the pattern was found to increase with increasing bias
voltage linearly at first, but to saturate at $\sim$8 nm where the
bias voltage reached $\sim$7 V. Bias voltages up to 12 V were
applied. In our experiment, the line depth increases in a stepwise
manner with increasing bias voltage, and the tip quality strongly
influence the etching results. At bias voltages above 3.6 V, mounds
or debris were formed on the surfaces and multi-tips were created
during etching, indicating that the Pt/Ir tip deteriorated above 3.6
V. This large different in upper limit for the applied voltage can
be explained by the sharper tips applied in the present work.
Sharper tips result in a stronger electric field formed between the
tip and the sample at lower voltages as compared blunter AFM tips.
At negative tip bias voltages, we observed both mounds and etched
lines appeared alternatively on LSMO surface, indicating the
evaporation from both tip and sample.

\begin{figure}[h!]
\begin{center}
  \includegraphics[width=8cm]{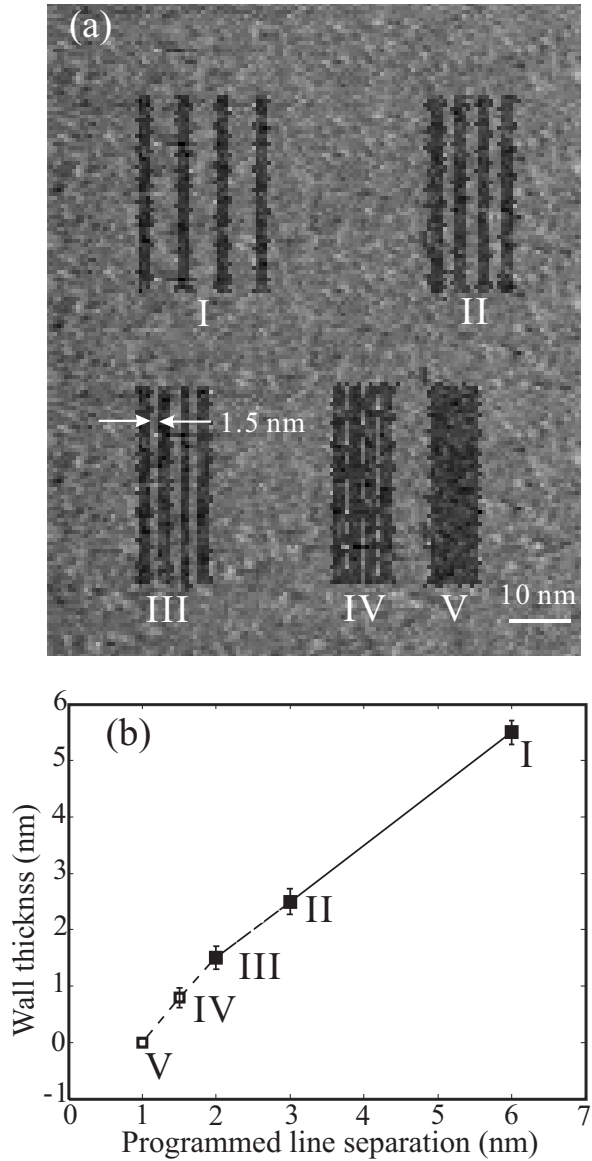}\\
  \caption{(a) Wall structures with minimum sizes of 1.5 nm, as pointed out by arrows. The etching
  parameters were 3.2 V bias voltage, 60 pA tunneling
  current, 1000 nm/s scanspeed and 100 scan repetitions per line.
 (b)Wall thickness as a function of programmed line
  separation. the solid line represents continuous walls, whereas the dashed line represents non-continuous and collapsed walls. }
  \label{fig:resolution}
\end{center}
\end{figure}

The different line depth values obtained in the line etching
experiments suggest that controlled etching by STM can be performed
in LSMO thin film surfaces. To determine the minimum wall thickness
which can be obtained between two consecutive etched lines, tests of
4-line patterns with decreasing separation were performed. Figure
\ref{fig:resolution}(a) shows five sets of etched lines with
different wall thicknesses. Figure \ref{fig:resolution}(b) is a plot
of wall thickness against programmed line separation. The offset
between programmed line separation and wall thickness is $\sim$0.5
nm. For programmed line separations 6 nm (I), 3 nm (II) and 2 nm
(III), the lines are separated by continuous walls. When the line
separation is 1.5 nm (IV), the walls are not continuous. Finally at
a line separation of 1 nm (V), the walls collapse, forming a
homogeneous etched area. Thus the minimum line width with continuous
walls from these experiments was determined to be 1.5 nm. The
structures in both figure \ref{fig:resolution} (a) was etched using
a bias voltage of 3.2 V, tunneling current of 60 pA, 1000 nm/s scan
speed and 100 scan repetitions per line. Most SPM-based lithography
techniques are able to create patterns with spatial resolution on
the order of 10 nm \cite{Tseng2005}. In their experiments performing
AFM lithography on LBMO thin films, Li et al \cite{Li2005} reported
both the line width and intervals are less than 50 nm. For other
established techniques, such as photolithography \cite{JaegerBook},
the minimum feature size is 30 nm, for focus ion beam milling, 30 nm
and for electron beam lithography \cite{Khoury1996,Tseng2003}, 10
nm.

\begin{figure}[h!]
\begin{center}
  \includegraphics[width=8cm]{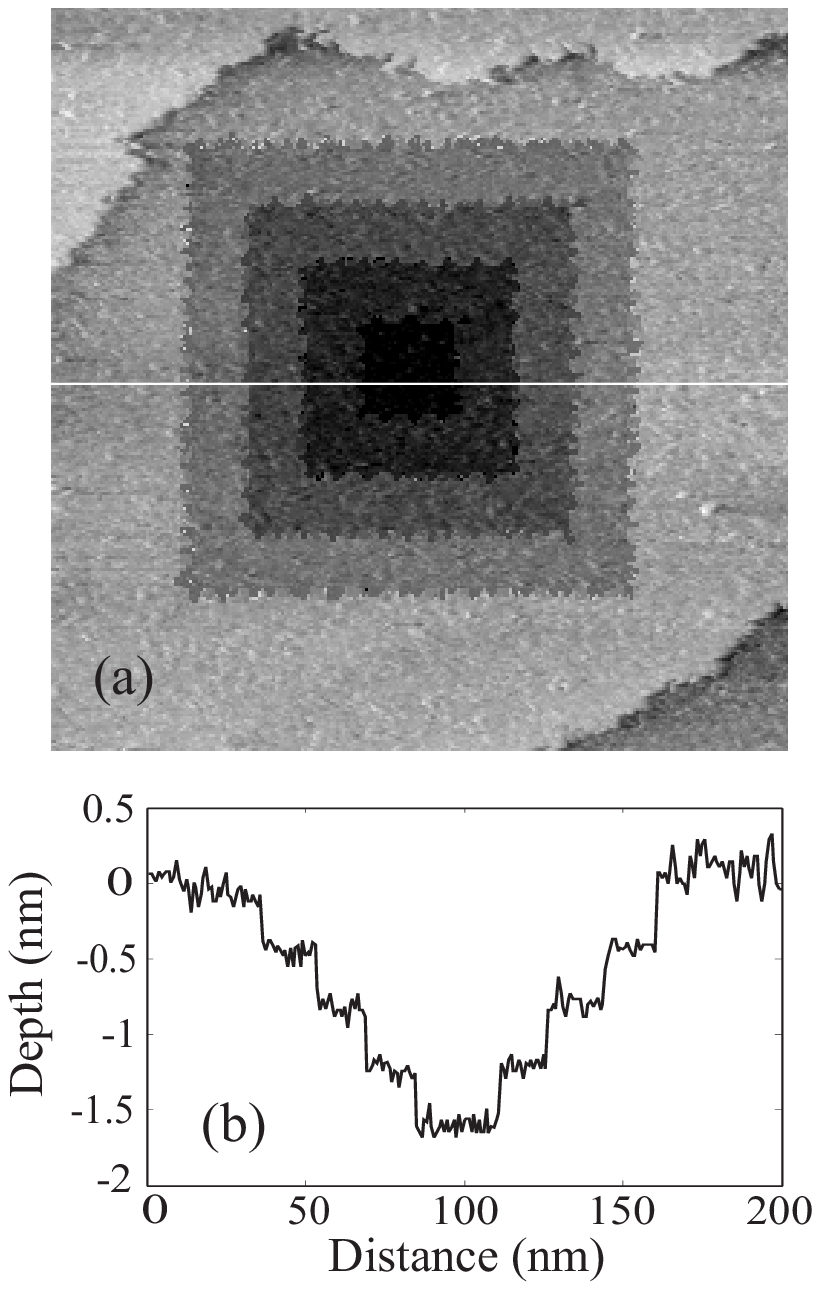}\\
  \caption{(a)Inverse-pyramid structures created by applying 3.2V bias voltage with setpoint tunneling current of
  60pA, scanspeed of 1000nm/s and 100 scan repetitions per line. (b)Depth profiles along  the mark lines in (a), showing
  step sizes of 0.4nm. The image area is 200nm$\times$200nm}
 \label{fig:inversepyramid}
\end{center}
\end{figure}

To demonstrate the capabilities of STM etching in LSMO we etched a
nanoscale architecture:  an inverse pyramid. It was constructed by
combining etched lines. Figure \ref{fig:inversepyramid} (a) shows
images of the inverse pyramid.  The etching parameters were the bias
voltages 3.2 V, 60 pA tunneling current, 1000 nm/s scan speed and
100 scan repetitions per line were. Figure
\ref{fig:inversepyramid}(b) display depth profiles along the marked
lines in figure \ref{fig:inversepyramid}(a). The step heights in
figure \ref{fig:inversepyramid}(a) was measured to be $\sim$0.4 nm,
which is consistent with the results of line etching in figure
\ref{fig:depthaVsvoltage}.

From both figure \ref{fig:inversepyramid} (a) we observed clearly
the zigzag patterns on the edges of structures. In order to
investigate the influence of in-plane crystallographic anisotropy on
this pattern, we conducted lithographic experiments through rotating
sample by 0$^{\circ}$, 45$^{\circ}$ and 90$^{\circ}$ relative to the
tip scanning directions. The results indicate that there are no
distinct differences in the etching results in these cases. In fact,
the periodic interval of protrudes along both the vertical and
horizontal edges is about 8 nm, hinting that this pattern could
result from the interval fashion of tip movement during etching
process.

\section{Conclusion}

In summary, controlled STM lithography has been performed on
perovskite manganite LSMO thin film surfaces. The line depth is
found to increase with increasing bias voltage and with decreasing
scan speed at a given number of scan repetitions per line. It is
observed that most values of average line depth are integral
multiples of c-axis constant of LSMO unit cell. Based on
line-etching experiments, we successfully created inverse-pyramid
nanoscale structures. Furthermore, $\sim$1.5 nm thick walls were
achieved as the smallest structures from STM lithography on LSMO
thin film surfaces to date.

\end{document}